\documentclass[twocolumn,showpacs,preprintnumbers,amsmath,amssymb]{revtex4}

\usepackage{graphicx}
\usepackage{dcolumn}
\usepackage{bm}

\newcommand{\bq}{\begin{equation}}
\newcommand{\eq}{\end{equation}}
\newcommand{\bqa}{\begin{eqnarray}}
\newcommand{\eqa}{\end{eqnarray}}
\newcommand{\nn}{\nonumber \\}

\def\be     {\begin{equation}}
\def\ee     {\end{equation}}
\def\bea        {\begin{eqnarray}}
\def\eea        {\end{eqnarray}}
\def\bnn    {\begin{eqnarray*}}
\def\enn    {\end{eqnarray*}}

\begin{document}

\title{An effective Lagrangian for the continuous transition in an extended Kondo lattice model}
\author{Ki-Seok Kim}
\affiliation{ School of Physics, Korea Institute for Advanced
Study, Seoul 130-012, Korea }
\date{\today}

\begin{abstract}
We propose an effective Lagrangian for the continuous transition
from the heavy fermion metal to the antiferromagnetic metal in an
extended Kondo lattice model. Based on the slave-boson
representation we introduce an additional new order parameter
associated with difference of the chemical potential between
conduction electrons $c_{i\sigma}$ and local spinons
$f_{i\sigma}$. This order parameter allows pseudospin construction
$T_{ix} = \frac{1}{2}\langle{c}_{i\alpha}^{\dagger}f_{i\alpha} +
f_{i\alpha}^{\dagger}c_{i\alpha}\rangle$, $T_{iy} = -
\frac{i}{2}\langle{c}_{i\alpha}^{\dagger}f_{i\alpha} -
f_{i\alpha}^{\dagger}c_{i\alpha}\rangle$, and $T_{iz} =
\frac{1}{2}\langle{c}_{i\alpha}^{\dagger}c_{i\alpha} -
f_{i\alpha}^{\dagger}f_{i\alpha}\rangle$, where $T_{i\pm} = T_{ix}
\pm iT_{iy}$ corresponds to the usual hybridization order
parameter in the slave-boson representation of the Kondo lattice
model. The resulting effective action is shown to be an
anisotropic pseudospin model with a Landau damping term for the
screened-unscreened (XY$-$Ising) phase transition. To describe the
emergence of antiferromagnetic order in the unscreened (Ising)
phase, we phenomenologically introduce the antiferromagnetic
Heisenberg model for the localized spins, where the effective
coupling strength is given by $J_{eff} =
J|\langle{T}_{iz}\rangle|^{2}$. This ad-hoc construction allows
the continuous transition from the heavy fermion phase to the
antiferromagnetic phase because breakdown of Kondo screening
($\langle{T}_{i\pm}\rangle = 0$ and $\langle{T}_{iz}\rangle \not=
0$) causes effective exchange interactions between unscreened
local moments.
\end{abstract}


\maketitle

\section{Introduction}

In the Landau-Ginzburg-Wilson (LGW) theoretical framework a
continuous phase transition is generically forbidden between two
different orders characterized by different symmetry breaking if
multi-critical points resulting from fine tuning of the couplings
in the LGW theory are not taken into account. Recently, Berry
phase mechanism was proposed for generic continuous transitions in
quantum antiferromagnets and Bose-Mott
insulators.\cite{Senthil_AF_DQCP,Kim_AF_DQCP,Tanaka,Balents_SC,Kim_SC}
If the disorder parameter in one phase transforms as the order
parameter in the other phase, the continuous transition between
the two phases is naturally expected to occur. However, this
situation does not happen generally. Condensation of skyrmions in
the antiferromagnetic phase or vortices in the superconducting
phase kills the ordered phase. The resulting disordered state
would be a paramagnetic phase or Bose-Mott insulating phase,
preserving all possible symmetries. But, the presence of Berry
phase can break the symmetries associated with lattice
translations or rotations since the Berry phase plays the role of
effective potentials in the disorder parameters, causing the
lattice-symmetry breaking. Critical field theories were also
proposed, and in some cases "exotic" quantum criticality is
expected to appear, described by gauge-interacting deconfined
particles carrying fractional quantum numbers.

Some heavy fermion systems also exhibit the LGW forbidden
continuous transition from the heavy fermion metal to the
antiferromagnetic metal, where the heavy fermion phase is
understood by hybridization between conduction electrons and local
moments while the antiferromagnetic state is described by the
magnetic long range order of localized spins.\cite{Kondo_Review}
The former is associated with an internal U(1) gauge symmetry like
superconductivity, and the latter, the spin rotational symmetry.
In such heavy fermion systems the Berry phase mechanism of the
magnetic and superconducting systems does not seem to work because
the disorder parameter in one phase does not transform as the
order parameter in the other phase even in the presence of Berry
phase.

Unfortunately, the mechanism of the continuous transition is still
far from consensus. In the slave-boson formulation the heavy
fermion phase naturally appears via the hybridization order
parameter while the antiferromagnetic phase arises from the
antiferromagnetic order parameter of localized
spins.\cite{Senthil_Kondo,Kim_Kondo1} Because these two order
parameters are associated with different symmetry breaking, the
continuous transition is generally forbidden in the LGW framework,
as discussed above. In the slave-fermion representation the
antiferromagnetic order naturally arises via the condensation of
bosonic spinons while the heavy fermion phase is described by new
emergent spinless fermions instead of some bosonic order
parameters.\cite{Pepin_Kondo,Kim_Kondo2} But, the strong Kondo
coupling regime is not easy to explain in this framework. On the
other hand, in the Bose-Fermi Kondo model the continuous
transition is expected to arise in low dimensions since
divergences of the local and global spin susceptibilities should
occur at the same time.\cite{Si} But, in three dimensions the
divergences in both spin susceptibilities would not happen
simultaneously.

In the present paper we revisit the slave-boson formulation of the
Kondo lattice model, and propose an effective Lagrangian for the
continuous transition from the heavy fermion metal to the
antiferromagnetic metal. We introduce an additional new order
parameter associated with difference of the chemical potential
between conduction electrons $c_{i\sigma}$ and local spinons
$f_{i\sigma}$, extending the usual slave-boson construction. This
order parameter allows pseudospin construction $T_{ix} =
\frac{1}{2}\langle{c}_{i\alpha}^{\dagger}f_{i\alpha} +
f_{i\alpha}^{\dagger}c_{i\alpha}\rangle$, $T_{iy} = -
\frac{i}{2}\langle{c}_{i\alpha}^{\dagger}f_{i\alpha} -
f_{i\alpha}^{\dagger}c_{i\alpha}\rangle$, and $T_{iz} =
\frac{1}{2}\langle{c}_{i\alpha}^{\dagger}c_{i\alpha} -
f_{i\alpha}^{\dagger}f_{i\alpha}\rangle$, where $T_{i\pm} = T_{ix}
\pm iT_{iy}$ corresponds to the usual hybridization order
parameter in the slave-boson representation of the Kondo lattice
model. The resulting effective action is shown to be an
anisotropic pseudospin model with a Landau damping term for the
screened-unscreened (XY$-$Ising) phase transition. To describe the
emergence of antiferromagnetic order in the unscreened (Ising)
phase, we phenomenologically introduce the antiferromagnetic
Heisenberg model for the localized spins, where the effective
coupling strength is given by $J_{eff} =
J|\langle{T}_{iz}\rangle|^{2}$. This ad-hoc construction allows
the continuous transition from the heavy fermion phase to the
antiferromagnetic phase because breakdown of Kondo screening
($\langle{T}_{i\pm}\rangle = 0$ and $\langle{T}_{iz}\rangle \not=
0$) causes effective exchange interactions between unscreened
local moments. We argue that our effective theory results in
discontinuous volume change of the Fermi surface across the
transition. Furthermore, we propose a critical field theory, and
discuss critical behaviors near the quantum critical point.

\section{Discussion on results}

We start from discussion of results in order to get a clear
physical picture. The effective action is proposed to be \bqa &&
S_{eff} = \frac{1}{\beta}\sum_{q,\Omega}(\xi_{\bot}^{-2} + q^{2} +
\gamma\frac{|\Omega|}{q})T_{+}(q,i\Omega)T_{-}(q,i\Omega) \nn && +
\frac{1}{\beta}\sum_{q,\Omega}(\xi_{\|}^{-2} + q^{2} +
\gamma\frac{|\Omega|}{q})T_{z}(q,i\Omega)T_{z}(-q,-i\Omega) \nn &&
+J|\langle{T}_{z}\rangle|^{2}\sum_{ij}\mathbf{S}_{i}\cdot\mathbf{S}_{j}
. \eqa Here $T_{i+} = T_{ix} + iT_{iy}$, $T_{i-} = T_{ix} -
iT_{iy}$, and $T_{iz}$ are the pseudospin fields, considered as
\bqa && T_{ix} = \frac{1}{2}(c_{i\alpha}^{\dagger}f_{i\alpha} +
f_{i\alpha}^{\dagger}c_{i\alpha}) , \nn && T_{iy} = -
\frac{i}{2}(c_{i\alpha}^{\dagger}f_{i\alpha} -
f_{i\alpha}^{\dagger}c_{i\alpha}) , \nn && T_{iz} =
\frac{1}{2}(c_{i\alpha}^{\dagger}c_{i\alpha} -
f_{i\alpha}^{\dagger}f_{i\alpha}) , \nonumber \eqa and
$\mathbf{S}_{i}$ the local spin fields, given by \bqa &&
\mathbf{S}_{i} =
\frac{1}{2}f_{i\alpha}^{\dagger}\vec{\sigma}_{\alpha\beta}f_{i\beta}
, \nonumber \eqa where $c_{i\alpha}$ is a conduction electron, and
$f_{i\alpha}$ a spinon for localized spins in the slave-boson
representation. One can find that the expectation value of the
pseudospin raising operator given by $\langle{T}_{i+}\rangle =
\langle{c}_{i\alpha}^{\dagger}f_{i\alpha}\rangle$ is nothing but
the hybridization order parameter in the slave boson
formulation.\cite{Senthil_Kondo,Kim_Kondo1} Thus, introduction of
the $z-$component pseudospin variable naturally extends the
slave-boson formulation. The $T_{iz}$ operator represents
difference of the chemical potential between the conduction
electrons and local spinons, considered to be another measure for
the Kondo screening. It is important to notice that introduction
of the $T_{iz}$ operator completes the commutation relation
$[T_{iz},T_{j\pm}] = \pm\delta_{ij}T_{i\pm}$, resulting in the
right uncertainty relation.

$\xi_{\bot}$ and $\xi_{\|}$ are the pseudospin correlation lengths
associated with the XY and Ising orders respectively, obtained to
be $\xi_{\bot} = \xi_{\|}$ in the SU(2) symmetric case.
Physically, $\xi_{\bot}$ corresponds to the Kondo screening
length, given by $\xi_{\bot}^{-2} \sim 1 - J_{K}D$, thus diverging
at the quantum critical point $J_{K}^{c} = 1/D$. Here $D$ is the
density of fermion states. We remark one crucial assumption
$\xi_{\|}^{-2}\cdot \xi_{\bot}^{-2} \leq 0$ in this paper. This
ad-hoc construction allows the continuous transition between the
XY and Ising orders of the pseudospin fields through the quantum
critical point $J_{K} = J_{K}^{c}$ satisfying $\xi_{\|}^{-2},
\xi_{\bot}^{-2} \rightarrow 0$.

In the case of $J_{K} > J_{K}^{c}$ ($\xi_{\|}^{-2} > 0$ and
$\xi_{\bot}^{-2} < 0$) the XY order appears to describe the heavy
fermion phase, characterized by $\langle{T}_{i+}\rangle =
\langle{c}_{i\alpha}^{\dagger}f_{i\alpha}\rangle \not= 0$ and
$\langle{T}_{iz}\rangle = \langle c_{i\alpha}^{\dagger}c_{i\alpha}
- f_{i\alpha}^{\dagger}f_{i\alpha} \rangle = 0$. This means that
there exist no unscreened local magnetic moments owing to the
Kondo hybridization, and the vacuum expectation value of the
localized spins vanishes, $\langle\mathbf{S}_{i}\rangle = 0$. In
the case of $J_{K} < J_{K}^{c}$ ($\xi_{\|}^{-2} < 0$ and
$\xi_{\bot}^{-2} > 0$) the Ising order arises to imply incomplete
Kondo screening, thus the heavy fermion phase disappears,
described by $\langle{T}_{i\pm}\rangle = 0$ and
$\langle{T}_{iz}\rangle \not= 0$. As a result, unscreened local
magnetic moments occur due to the chemical potential difference,
causing effective exchange interactions
$J|\langle{T}_{z}\rangle|^{2}$ between the local moments. This
mechanism is reflected in the antiferromagnetic Heisenberg model
of localized spins, resulting in the antiferromagnetic long range
order $\langle\mathbf{S}_{i}\rangle \not= 0$ in the unscreened
(Ising) phase. Because the effective exchange interactions depend
on the vacuum expectation value of the $T_{iz}$ operator, the
antiferromagnetic order should start from the quantum critical
point. As a result, the second order transition from the heavy
fermion phase to the antiferromagnetic one is truly achieved.
Unfortunately, we failed the derivation of these effective
exchange couplings. This ad-hoc construction is another important
assumption in this paper.

The XY and Ising orders become degenerate at the quantum critical
point because of $\xi_{\|}^{-2}, \xi_{\bot}^{-2} \rightarrow 0$,
indicating the emergence of a new symmetry that is absent away
from the quantum critical point. The emerging symmetry is SU(2)
pseudospin symmetry larger than U(1) in the XY phase and Z(2) in
the Ising state. This enlarged symmetry plays an important role in
critical dynamics of pseudospins with the frequency-dependent
terms in Eq. (1) known as Landau damping terms with a damping
coefficient $\gamma$, resulting from low energy fermion
excitations near a Fermi surface. This will be discussed later.

It should be noted that our effective action exhibits a
discontinuous change in the Fermi surface volume. In the Kondo
screened phase a large Fermi surface is expected owing to the
Kondo hybridization $\langle{T}_{\pm}\rangle \not= 0$. On the
other hand, in the decoupled phase a small Fermi surface would
appear due to $\langle{T}_{\pm}\rangle = 0$. Although the XY to
Ising transition is continuous, discontinuity in the volume change
of the Fermi surface should be observed right at the critical
point since the large Fermi surface is sustained as long as the
Kondo hybridization exists.

A schematic phase diagram is shown in Fig. 1, where HF denotes the
heavy fermion phase, AF the antiferromagnetic phase, and QC the
quantum critical regime. Since HF and AF are described by the XY
and Ising orders in terms of the pseudospin fields approximately,
one might expect that their finite temperature transitions fall
into the Berezinskii-Kosterlitz-Thouless (BKT) transition in two
dimensions and Ising one, respectively. Precisely speaking, this
guess is not correct because the Landau damping term modifies the
transition nature apparently, as will be discussed later. Because
there is no finite temperature transition in the SU(2) symmetric
model owing to the Mermin-Wagner-Hohenberg-Coleman (MWHC) theorem,
both the XY and Ising transition lines meet and vanish at the
SU(2) symmetric quantum critical point, i.e. $T_{XY} = T_{Ising}
\rightarrow 0$.

\begin{figure}
\includegraphics[width=8cm]{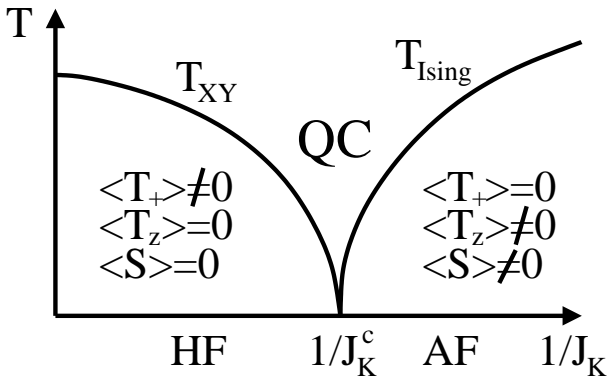}
\caption{\label{Fig. 1} A schematic phase diagram in the effective
action of the Kondo lattice model}
\end{figure}

\section{Derivation of the effective Lagrangian}

We derive an effective pseudospin action from the Kondo lattice
model \bqa && H = H_{c} + H_{m} + H_{K} , \nn && H_{c} =
-t\sum_{ij} c_{i\alpha}^{\dagger}c_{j\alpha} , \nn && H_{m} =
J\sum_{ij}\mathbf{S}_{i}\cdot\mathbf{S}_{j} , \nn && H_{K} =
J_{K}\sum_{i}\mathbf{S}_{i}\cdot{c}^{\dagger}_{i\alpha}\vec{\sigma}_{\alpha\beta}{c}_{i\beta}
, \eqa where $J_{K}$ is the Kondo coupling between the conduction
electrons and local spins, and $J$ the antiferromagnetic coupling
between the local spins.

Using the slave-boson representation, one can obtain a one-body
effective Lagrangian in terms of the conduction electrons and
spinons coupled to order parameters\cite{Senthil_Kondo,Kim_Kondo1}
\bqa && Z =
\int{Dc_{\alpha}}{Df_{\alpha}}{Db}e^{-\int_{0}^{\beta}{d\tau} L} ,
\nn && L = \sum_{i}c_{i\alpha}^{\dagger}(\partial_{\tau} -
\mu_{c})c_{i\alpha} - t\sum_{ij}c_{i\alpha}^{\dagger}c_{j\alpha}
\nn && + \sum_{i}f_{i\alpha}^{\dagger}(\partial_{\tau} -
\mu_{f})f_{i\alpha} -
\chi_{0}\sum_{ij}f^{\dagger}_{i\alpha}f_{j\alpha} +
\sum_{ij}\frac{\chi_{0}^{2}}{J} \nn && -
\sum_{i}(b^{\dagger}_{i}c_{i\alpha}^{\dagger}f_{i\alpha} + H.c.) +
\sum_{i}\frac{|b_i|^2}{J_K} . \eqa Here $\mu_{c}$ is the chemical
potential of conduction electrons, and $\mu_{f}$ that of local
spinons. $b_{i}$ is called holon, representing hybridization
between the conduction electron and spinon. $\chi_{0}$ is the
hopping order parameter of the spinons, assumed to be spatially
homogeneous. Mean field values of the order parameters are given
by \bqa && b_{i} = J_{K}<c_{i\alpha}^{\dagger}f_{i\alpha}> , \nn
&& \chi_{0} = \frac{J}{2}<f_{i\alpha}^{\dagger}f_{j\alpha}> .
\nonumber \eqa Here we consider the case of half filling for the
spinons, $1 = <f_{i\alpha}^{\dagger}f_{i\alpha}>$.

As mentioned in the introduction, the slave-boson effective
Lagrangian is difficult to describe the continuous transition to
the antiferromagnetic long range order in the mean field level. In
this respect an extension of Eq. (3) is indispensable. It is
convenient to express Eq. (3) in terms of the "Nambu" spinor \bqa
&& \psi_{in\alpha} = \left(\begin{array}{c} \psi_{i1\alpha}
\\ \psi_{i2\alpha} \end{array} \right) = \left(\begin{array}{c} c_{i\alpha}
\\ f_{i\alpha} \end{array} \right) .
\eqa Inserting this into Eq. (3), we obtain \bqa && L_{eff} =
\sum_{i}\psi_{in\alpha}^{\dagger}(\partial_{\tau} -
\mu_{n})\psi_{in\alpha} -
\sum_{ij}t_{n}\psi_{in\alpha}^{\dagger}\psi_{jn\alpha} \nn &&
-\sum_{i}\psi_{in\alpha}^{\dagger}({T}_{i+}{\tau}_{nm}^{+} +
{T}_{i-}{\tau}_{nm}^{-} )\psi_{im\alpha} +
\sum_{i}\frac{{T}_{i+}T_{i-}}{J_{K}} \eqa with $\mu_{n} =
\left(\begin{array}{c} \mu_{c} \\ \mu_{f}
\end{array} \right)$ and $t_{n} = \left(\begin{array}{c} t_{c} \\
\chi_{0} \end{array} \right)$. Here $T_{i+}$ corresponds to the
holon field $b_{i}$.

The $T_{i+}$ and $T_{i-}$ operators can be considered as the
rasing and lowering operators of the pseudospin variable if we
introduce the $z-$component pseudospin field $T_{iz}$. Consider
the following term \bqa && L_{z} = -
\frac{U_{K}}{4}\sum_{i}(c_{i\alpha}^{\dagger}c_{i\alpha} -
f_{i\alpha}^{\dagger}f_{i\alpha})^{2} \nn && = -
U_{K}\sum_{i}(\psi_{in\alpha}^{\dagger}{\tau}^{z}_{nm}
\psi_{im\alpha})^{2} . \eqa These local interactions compete with
the Kondo couplings because this term favors unscreened local
moments. Performing the Hubbard-Stratonovich transformation, and
adding the resulting Lagrangian \bqa && L_{z} = - \sum_{i}
\psi_{in\alpha}^{\dagger}{T}_{iz}{\tau}^{z}_{nm} \psi_{im\alpha} +
\sum_{i}\frac{{T}_{iz}^{2}}{4U_{K}} \nonumber \eqa into Eq. (5),
where $T_{iz}$ is the $z-$component pseudospin variable, the
effective Lagrangian Eq. (5) can be naturally generalized into
\bqa && L_{eff} =
\sum_{i}\psi_{in\alpha}^{\dagger}(\partial_{\tau} -
\mu_{n})\psi_{in\alpha} -
\sum_{ij}t_{n}\psi_{in\alpha}^{\dagger}\psi_{jn\alpha} \nn &&
-\sum_{i}\psi_{in\alpha}^{\dagger}\mathbf{T}_{i}\cdot\vec{\tau}_{nm}
\psi_{im\alpha} + \sum_{i}\Bigl(\frac{{T}_{i+}T_{i-}}{4J_{K}} +
\frac{{T}_{iz}^{2}}{4U_{K}}\Bigr) , \nn \eqa where $T_{ix(y)}$ was
replaced with $(1/2)T_{ix(y)}$. Here the pseudospin variable is
obtained to be in the mean field level \bqa &&
\frac{T_{ix}}{4J_{K}} =
\frac{1}{2}\langle{c}_{i\alpha}^{\dagger}f_{i\alpha} +
f_{i\alpha}^{\dagger}c_{i\alpha}\rangle = \frac{1}{2}
\langle\psi_{in\alpha}^{\dagger}\tau^{x}_{nm}\psi_{im\alpha}\rangle
, \nn && \frac{T_{iy}}{4J_{K}} = -
\frac{i}{2}\langle{c}_{i\alpha}^{\dagger}f_{i\alpha} -
f_{i\alpha}^{\dagger}c_{i\alpha}\rangle = \frac{1}{2}
\langle\psi_{in\alpha}^{\dagger}\tau^{y}_{nm}\psi_{im\alpha}\rangle
, \nn && \frac{T_{iz}}{4U_{K}} =
\frac{1}{2}\langle{c}_{i\alpha}^{\dagger}c_{i\alpha} -
f_{i\alpha}^{\dagger}f_{i\alpha}\rangle = \frac{1}{2}
\langle\psi_{in\alpha}^{\dagger}\tau^{z}_{nm}\psi_{im\alpha}\rangle
. \nonumber \eqa

The chemical potential term in Eq. (7) can be written as \bqa &&
-\mu_{n}\psi_{in\alpha}^{\dagger}\psi_{in\alpha} = -
\mu\psi_{in\alpha}^{\dagger}\psi_{in\alpha} -
\mu_{z}\psi_{in\alpha}^{\dagger}\tau_{nm}^{z}\psi_{im\alpha}
\nonumber \eqa with $\mu = (\mu_{c} + \mu_{f})/2$ and $\mu_{z} =
(\mu_{c} - \mu_{f})/2$. Shifting the $z-$component pseudospin as
$T_{iz} \rightarrow T_{iz} - \mu_{z}$, we obtain the effective
Lagrangian \bqa && L_{eff} =
\sum_{i}\psi_{in\alpha}^{\dagger}(\partial_{\tau} -
\mu)\psi_{in\alpha} -
\sum_{ij}t_{n}\psi_{in\alpha}^{\dagger}\psi_{jn\alpha} \nn &&
-\sum_{i}\psi_{in\alpha}^{\dagger}\mathbf{T}_{i}\cdot\vec{\tau}_{nm}
\psi_{im\alpha} + \sum_{i}\Bigl(\frac{{T}_{i+}T_{i-}}{4J_{K}} +
\frac{({T}_{iz}- \mu_{z})^{2}}{4U_{K}} \Bigr) . \nn \eqa It should
be noticed that the bare chemical potential of the conduction
electrons is the same as that of the local spinons. The $T_{iz}$
pseudospin field controls the chemical potential difference
between the conduction electrons and spinons, managing the Kondo
screening.

Integrating out the spinor fields in Eq. (8), and expanding the
resulting logarithmic action for the pseudospin fields, one can
obtain an effective Lagrangian in terms of the pseudospin
variables. There are two approaches; one is a strong coupling
approach valid in the limit of $t_{n} << |\mathbf{T}| \equiv
M_{K}$, and the other a weak coupling one justified in the limit
of $t_{n} >> M_{K}$, where $M_{K}$ is amplitude of the pseudospin.

\subsection{Strong coupling approach}

In the strong coupling limit we consider the $CP^{1}$
representation for the pseudospin variable \bqa &&
\mathbf{T}_{i}\cdot\vec{\tau}_{nm} = M_{K}
U_{inp}\tau_{pq}^{z}U_{iqm}^{\dagger} , \nn && U_{inm} = \left( \begin{array}{cc} z_{i1} & -z_{i2}^{\dagger} \\
z_{i2} & z_{i1}^{\dagger} \end{array} \right) , \eqa where
$U_{inm}$ is a SU(2) matrix field in terms of a complex boson
field $z_{in}$ with the pseudospin index $n = 1, 2$, and $M_{K}$
amplitude of the pseudospin. Inserting Eq. (9) into Eq. (8), and
performing the gauge transformation \bqa && \psi_{in\alpha} =
U_{inm}\Psi_{im\alpha} , \eqa we obtain \bqa && L_{eff} =
\sum_{i}\Psi_{ik\alpha}^{\dagger}([\partial_{\tau} -
\mu]\delta_{km} +
U^{\dagger}_{ikn}\partial_{\tau}U_{inm})\Psi_{im\alpha} \nn && -
\sum_{ij}t_{n}\Psi_{ik\alpha}^{\dagger}U_{ikn}^{\dagger}U_{jnm}\Psi_{jm\alpha}
-\sum_{i}M_{K}\Psi_{ik\alpha}^{\dagger}\tau_{km}^{z}\Psi_{im\alpha}
. \nn \eqa

Integrating out the spinor fields $\Psi_{i}$, and expanding the
resulting logarithmic term for
$U_{i}^{\dagger}\partial_{\tau}U_{i}$ and $U_{i}^{\dagger}U_{j}$,
we obtain \bqa && S_{eff} = - \mathbf{tr}ln\Bigl[(\partial_{\tau}
- \mu)\delta_{km} - M_{K}\tau^{z}_{km} \nn && +
U_{ikn}^{\dagger}\partial_{\tau}U_{inm} -
t_{n}U_{ikn}^{\dagger}U_{jnm}\Bigr] \nn && \approx
\sum_{i}\mathbf{tr}[G_{0}(U_{i}^{\dagger}\partial_{\tau}U_{i})]
\nn && +
\frac{1}{2}\sum_{i}\mathbf{tr}_{j}[G_{0}U_{i}^{\dagger}\mathbf{t}U_{j}G_{0}U_{j}^{\dagger}\mathbf{t}U_{i}]
, \eqa where $G_{0} = - (\partial_{\tau} - \mu -
M_{K}\tau_{z})^{-1}$ is the single particle propagator. Notice
that the spinor propagator depends on only time or frequency, not
position or momentum, implying that the kinetic energy of
electrons and spinons is ignored owing to strong hybridization.
This means that the strong coupling expansion justified in the
limit of $t << M_{K}$ can be applied to an insulating phase of
electrons and spinons. In other words, the assumed ground state of
spinors is an insulating phase in this expansion scheme. This will
be more clarified later. The first term leads to Berry phase while
the second results in an exchange interaction term, both well
evaluated in Refs. \cite{Kim_SC,Nagaosa}. One difference from the
previous studies\cite{Kim_SC,Nagaosa} is the presence of the
chemical potential term. But, this does not result in any
important modification.

The resulting effective action is obtained to be \bqa && S_{eff} =
iS\sum_{i}\omega(\{\mathbf{T}_{i}(\tau)\}) +
\int_{0}^{\beta}{d\tau} H_{eff} , \nn && H_{eff} =
I_{K}\sum_{ij}(T_{ix}T_{jx} + T_{iy}T_{jy}) +
V_{K}\sum_{ij}T_{iz}T_{jz} , \nn \eqa where $S$ is the pseudospin
value, given by $S = 1/2$. It is interesting that the effective
Hamiltonian for the Kondo screening is given by the
antiferromagnetic pseudospin model in the strong coupling limit.
The first term in $S_{eff}$ is a Berry phase term of the
pseudospin field. $I_{K}$ and $V_{K}$ are the XY and Ising
exchange couplings, given by $I_{K} = V_{K} = 2t^2/M_{K}$ for the
SU(2) symmetric case.\cite{Kim_SC,Nagaosa} Here the anisotropy in
the hopping integrals is neglected for simplicity, discussed later
in more detail.

Eq. (13) shows a continuous transition from the XY order to the
Ising one at the quantum critical point $V_{K}/I_{K} = 1$. If
$1/J_{K}$ is replaced with $V_{K}/I_{K}$ in Fig. 1, the
antiferromagnetic pseudospin model exhibits a similar phase
diagram with Fig. 1.\cite{Zhang} In the limit of $V_{K}/I_{K} <<
1$ a Kondo insulator is expected while in the opposite limit an
antiferromagnetic insulator would result. The finite temperature
transitions are simply given by the BKT transition in the two
dimensional XY phase of $V_{K}/I_{K} < 1$ and the Ising one in the
Ising state of $V_{K}/I_{K} > 1$, respectively. These transition
lines should meet and vanish at the quantum critical point owing
to the enlarged SU(2) pseudospin symmetry (MWHC theorem).

\subsection{Weak coupling approach}

In the weak coupling approach valid in the $t >> M_{K}$ limit, one
obtains the following effective action after integrating out the
spinor fields $\psi_{in\alpha}$ in Eq. (8) \bqa && S_{eff} = -
\mathbf{tr}ln\Bigl[
\partial_{\tau} - \mu - t_{ij} - \mathbf{T}_{i}\cdot\tau \Bigr] .
\eqa Note the difference between Eq. (12) and Eq. (14). The main
difference is that the spinor propagator in Eq. (14) includes the
momentum (position) dependence, implying that the assumed ground
state is a noninteracting Fermi gas, thus metal. In this respect
the weak coupling approach can be applied to a metallic phase.

Expanding the logarithmic action for the pseudospin field, one
finds \bqa && S_{eff} =
\frac{1}{\beta}\sum_{q,\Omega}\Bigl(\chi^{-1}_{\bot}(q,i\Omega)T_{+}(q,i\Omega)T_{-}(q,i\Omega)
\nn && +
\chi^{-1}_{\|}(q,i\Omega)T_{z}(q,i\Omega)T_{z}(-q,-i\Omega) \Bigr)
\nn && +
u\int_{0}^{\beta}{d\tau}{d^2r}[\mathbf{T}(r,\tau)\cdot\mathbf{T}(r,\tau)]^{2}
. \eqa Here $\chi^{-1}_{\bot}(q,i\Omega)$ and
$\chi^{-1}_{\|}(q,i\Omega)$ are transverse and longitudinal
pseudospin susceptibilities, and $u$ a local interaction parameter
of the pseudospin fields. The pseudospin susceptibility is
evaluated in the noninteracting fermion ensemble, thus given
by\cite{Nagaosa,FM_QCP} \bqa && \chi^{-1}_{\bot}(q,i\Omega) =
\xi_{\bot}^{-2} + q^{2} + \gamma|\Omega|/q , \nn &&
\chi^{-1}_{\|}(q,i\Omega) = \xi_{\|}^{-2} + q^{2} +
\gamma|\Omega|/q . \eqa Here $\xi_{\bot}$ is the Kondo screening
length, obtained to be $\xi_{\bot}^{-2} \sim 1 - J_{K}D$, where
$D$ is the density of states of spinors. It diverges at the
quantum critical point $J_{K}^{c} = 1/D$. $\xi_{\|}$ is another
pseudospin correlation length associated with the Ising order. In
the SU(2) symmetric case one obtains $\xi_{\bot} = \xi_{\|}$. As
mentioned before, the following relation $\xi_{\|}^{-2}\cdot
\xi_{\bot}^{-2} \leq 0$ is an important assumption in this paper.
This assumption is parallel to the pseudospin anisotropy in Eq.
(13). This ad-hoc construction allows the continuous transition
between the XY and Ising orders of the pseudospin fields. The
presence of the Landau damping term confirms that this approach is
based on the metallic phase since it originates from low energy
fermion excitations near the Fermi surface.

Eq. (15) is nothing but the Hertz-Millis theory for the magnetic
quantum phase transition in itinerant electron
systems.\cite{FM_QCP} Recently, it was claimed that the
Hertz-Millis theory may not be applied because the Landau
expansion in Eq. (14) can result in a non-analytic correction to
the spin susceptibility instead of the normal $q^{2}$ contribution
beyond the Eliashberg approximation.\cite{Non_Analytic_Correction}
In this paper we limit our consideration within the Eliashberg
theory, thus do not allow the non-analytic correction to the
pseudospin susceptibility.

The above effective action does not include spin degrees of
freedom, thus not allowing antiferromagnetic order. To describe
the antiferromagnetic transition, we introduce the
antiferromagnetic Heisenberg Hamiltonian for the localized spins
\bqa && H_{AF} = J_{eff}\sum_{ij}\mathbf{S}_{i}\cdot\mathbf{S}_{j}
, \eqa where $J_{eff}$ is an effective coupling strength. Because
the antiferromagnetic long range order should start from the
transition point $J_{K} = J_{K}^{c}$ associated with the Kondo
screening, it is natural to set $J_{eff} =
J|\langle{T}_{z}\rangle|^{2}$ with the bare coupling strength $J$.
Remember that the $T_{iz}$ operator leads to the difference of
chemical potential between the conduction electrons and local
spinons. In the case of $J_{K} < J_{K}^{c}$ the Kondo
hybridization vanishes, resulting in incomplete Kondo screening
owing to the chemical potential difference. As a result,
unscreened local magnetic moments appear, generating the effective
exchange interactions. The presence of local moments is a
necessary condition for the antiferromagnetic long range order.
Unfortunately, we cannot derive this effective coupling constant
as mentioned before. However, this simple construction results in
the continuous transition from the heavy fermion phase to the
antiferromagnetic one in a very simple way.

\subsection{Hubbard model for Kondo phenomena}

In this section we give an alternative view for the pseudospin
model as an effective action. Consider the following Lagrangian
\bqa && L_{eff} =
\sum_{i}\psi_{in\alpha}^{\dagger}(\partial_{\tau} -
\mu_{n})\psi_{in\alpha} -
\sum_{ij}t_{n}\psi_{in\alpha}^{\dagger}\psi_{jn\alpha} \nn && +
U\sum_{i}N_{i+}N_{i-} , \eqa where the spinor $\psi_{in\alpha}$,
the hopping integral $t_{n}$, and the chemical potential $\mu_{n}$
are defined in Eq. (4) and Eq. (5), respectively. The Hubbard
interaction consists of $N_{i+} =
\psi_{i1\alpha}^{\dagger}\psi_{i1\alpha} =
c_{i\alpha}^{\dagger}c_{i\alpha}$ and $N_{i-} =
\psi_{i2\alpha}^{\dagger}\psi_{i2\alpha} =
f_{i\alpha}^{\dagger}f_{i\alpha}$, indicating that local repulsive
interactions work between the conduction electrons and spinons. As
well known, the Hubbard interaction can be decomposed into
"charge" and "spin" channels in the following way \bqa &&
(\psi_{i1\alpha}^{\dagger}\psi_{i1\alpha})(\psi_{i2\alpha}^{\dagger}\psi_{i2\alpha})
\nn && = \frac{1}{4}[\psi_{in\alpha}^{\dagger}\psi_{in\alpha}]^{2}
- \frac{1}{4}[\psi_{in\alpha}^{\dagger}({\vec
\Omega}_{i}\cdot{\vec \tau})_{nm}\psi_{im\alpha}]^{2} , \eqa where
${\vec \Omega}_{i}$ is a unit spin vector.

Performing the Hubbard-Stratonovich transformation, we obtain \bqa
&& L_{eff} = \sum_{i}\psi_{in\alpha}^{\dagger}(\partial_{\tau} -
\mu_{n})\psi_{in\alpha} -
\sum_{ij}t_{n}\psi_{in\alpha}^{\dagger}\psi_{jn\alpha} \nn && +
\sum_{i}\Bigl(\frac{1}{U}\Delta_{in}^{2} -
i\Delta_{in}\psi_{in\alpha}^{\dagger}\psi_{in\alpha}\Bigr) \nn &&
+ \sum_{i}\Bigl(\frac{1}{U}\mathbf{M}_{i}^{2} -
\mathbf{M}_{i}\cdot\psi_{in\alpha}^{\dagger}{\vec
\tau}_{nm}\psi_{im\alpha}\Bigr) , \eqa where $\Delta_{in}$ is an
effective potential for total number density, and $\mathbf{M}_{i}$
an effective magnetic field for the pseudospin
$\psi_{in\alpha}^{\dagger}{\vec \tau}_{nm}\psi_{im\alpha}$,
corresponding to $\mathbf{T}_{i}$. Eq. (20) is essentially the
same as Eq. (8) if we introduce the pseudospin anisotropy and
ignore the charge channel. This implies that the Hubbard model Eq.
(18) can be considered as an effective model for the Kondo
phenomena.

The strong coupling limit $t_{n} << U$ in Eq. (18) results in the
antiferromagnetic Heisenberg model Eq. (13) in terms of the
pseudospin fields. Since the strong coupling approach is based on
an insulating phase of spinors, this approach can be applied to
the Kondo insulators, as discussed before. On the other hand, in
the weak coupling limit $t_{n} >> U$ Eq. (18) describes an
itinerant spinor system. Thus, both ferromagnetic and
antiferromagnetic pseudospin instabilities can occur near the
Fermi surface. In this paper we do not take into account Fermi
surface nesting, thus only consider a ferromagnetic pseudospin
state.

\section{Quantum critical point}

We investigate critical pseudospin dynamics near the quantum
critical point. The effective action would be written in terms of
the spinor field $\psi_{n\alpha}$ interacting via critical
ferromagnetic pseudospin fluctuations $\mathbf{T}_{i}$, given by
\bqa && S_{eff} =
\frac{1}{\beta}\sum_{q,\Omega}\Bigl(\chi^{-1}_{\bot}(q,i\Omega)T_{+}(q,i\Omega)T_{-}(q,i\Omega)
\nn && +
\chi^{-1}_{\|}(q,i\Omega)T_{z}(q,i\Omega)T_{z}(-q,-i\Omega) \Bigr)
+ u\int_{0}^{\beta}{d\tau}{d^2r}(\mathbf{T}\cdot\mathbf{T})^{2}
\nn && + \int_{0}^{\beta}{d\tau}{d^2r}\Bigl[
\psi_{n\alpha}^{\dagger}(\partial_{\tau} - \mu)\psi_{n\alpha} +
\frac{1}{2m_{n}}|\nabla\psi_{n\alpha}|^{2} \nn && -
\psi_{n\alpha}^{\dagger}\mathbf{T}\cdot\vec{\tau}_{nm}
\psi_{m\alpha} \Bigr] , \eqa where $m_{n} \sim t_{n}^{-1}$ is the
bare band mass of spinors. The antiferromagnetic interaction term
may be ignored since its coupling strength vanishes at the quantum
critical point. However, we admit that it is difficult to justify
the ignorance of antiferromagnetic spin fluctuations. In this
paper we investigate the role of critical Kondo fluctuations in
critical phenomena.

We conjecture that SU$_{p}$(2) pseudospin symmetry appears at the
critical point. Note that the SU$_{p}$(2) symmetry does not exist
in the microscopic model, the Kondo lattice Hamiltonian since the
chemical potential and hopping integral of conduction electrons
differ from those of spinons. At the quantum critical point the
chemical potential of spinons is the same as that of conduction
electrons. One uncertainty is about the hopping integral $t_{n}$
at the critical point. We conjecture that the hopping integral of
conduction electrons is the same as that of spinons at the quantum
critical point. Even if this conjecture is not correct, the
difference of hopping integrals between the conduction electrons
and local spinons would not cause severe differences for critical
phenomena, compared to those of the SU$_{p}$(2) symmetric case. As
a result, SO(4) $\sim$ SU$_{p}$(2)$\bigotimes$ SU$_{s}$(2)
symmetry is expected to emerge at the quantum critical point,
where SU$_{s}$(2) is the spin rotation symmetry. In terms of the
symmetry language the heavy fermion phase is understood as
breaking the U$_{p}$(1) symmetry but preserving the SU$_{s}$(2)
symmetry while the antiferromagnetic phase is associated with
breaking the Z$_{p}$(2) symmetry and the SU$_{s}$(2) symmetry.
Unfortunately, Eq. (21) cannot describe the SU$_{s}$(2) symmetry
breaking, as mentioned before. To summarize, the critical action
for the Kondo screening is the same as that describing the
ferromagnetic quantum criticality.

One important point for the ferromagnetic quantum
criticality\cite{FM_QCP} is the Landau damping term in Eq. (16).
The propagator of ferromagnetic pseudospin fluctuations is given
by $\chi_{\bot}(q, i\Omega) = \chi_{\|}(q, i\Omega) = 1/(q^{2} +
\gamma|\Omega|/q)$ at the quantum critical point $\xi_{\bot} =
\xi_{\|} \rightarrow \infty$. As shown in the pseudospin
susceptibility, the dispersion of ferromagnetic pseudospin
fluctuations is given by $\Omega \sim q^{3}$, thus the dynamical
critical exponent $z$ is obtained to be $z = 3$. Since the
effective dimensionality of the system is $D_{eff} = d + z$ with
$d$ spatial dimensions, it is given by $D_{eff} = 5$ in $d = 2$
and $D_{eff} = 6$ in $d = 3$. Considering that the upper critical
dimension is $D_{up} = 4$ in the O(3) vector model, the
ferromagnetic quantum critical point is above the upper critical
dimension, thus described by the Gaussian theory for pseudospin
fluctuations. It is valuable to note that the $z = 3$ Gaussian
critical theory is found in the context of U(1) spin liquid with a
Fermi surface.\cite{U1SL}

The spinor self-energy is given by \bqa && \Sigma(k,i\omega) =
\frac{1}{\beta}\sum_{\Omega}\int\frac{d^dq}{(2\pi)^{d}}G(k+q,i\omega+i\Omega)\chi_{\bot}(q,i\Omega)
, \nonumber \eqa where $G(k,i\omega) = 1/(i\omega - \mu -
\epsilon_{k})$ is the spinor propagator with its dispersion
$\epsilon_{k}$. Scattering with $z = 3$ critical fluctuations is
well known to give $\Sigma'' \sim T^{2/3}$ in $d = 2$ and
$\Sigma'' \sim T$ in $d = 3$,\cite{Kim_Kondo1} where $\Sigma''$ is
the imaginary part of the self-energy. Remember $\Sigma'' \sim
T^{2}$ in the Fermi liquid. In a similar expression the scattering
rate $\tau^{-1}$ associated with the dc conductivity is obtained
to be $\tau^{-1} \sim T^{4/3}$ in $d = 2$ and $\tau^{-1} \sim
T^{5/3}$ in $d = 3$.\cite{U1SL_Lee} The specific heat $C_{v}$ in
the quantum critical regime can be obtained from $C_{v} \sim
\partial^{2}F_{s}/\partial^{2}T$, where $F_{s}$ is the singular
contribution of the free energy due to Gaussian fluctuations of
the critical pseudospin modes, given by \bqa && F_{s} = \beta^{-1}
\sum_{\Omega}\int\frac{d^dq}{(2\pi)^{d}} ln\chi_{\bot}(q,i\Omega)
. \nonumber \eqa The specific heat is obtained to be $C_{v} \sim
T^{2/3}$ in $d = 2$ and $C_{v} \sim TlnT$ in $d =
3$.\cite{Kim_Kondo1} Thus, $C_{v}/T$ diverges in the $T
\rightarrow 0$ limit at the quantum critical point.

Since the contribution of spin excitations in local moments is not
taken into account in the present critical theory, we cannot
explain the $\Omega/T$ scaling with an anomalous exponent in the
magnetic susceptibility.\cite{Kondo_Review} This remains an
important future work in this direction.

\section{Summary}

Introducing the new order parameter associated with the chemical
potential difference between the conduction electrons and local
spinons, we proposed an effective theory in terms of the
pseudospin field for the continuous transition from the heavy
fermion phase to the antiferromagnetic phase in the slave-boson
representation of the extended Kondo lattice Hamiltonian. Our
pseudospin construction shows an abrupt change in the Fermi
surface volume across the transition. Furthermore, we argued that
the quantum critical point in the present critical theory ignoring
antiferromagnetic spin fluctuations of the localized spins is
understood in the context of the ferromagnetic quantum criticality
of itinerant electron systems.

K.-S. Kim thanks Dr. K. Park for helpful discussion.

\end{document}